\documentclass[conference]{IEEEtran}
\usepackage[latin9]{inputenc}
\usepackage{amsmath}
\usepackage{amssymb}
\usepackage{graphicx}
\usepackage{fancyhdr}
 
\makeatletter
\IEEEoverridecommandlockouts
\usepackage{cite}
\usepackage{amsfonts}\usepackage{algorithmic}
\usepackage{subfigure}
\usepackage{textcomp}
\usepackage{xcolor}
\def\BibTeX{{\rm B\kern-.05em{\sc i\kern-.025em b}\kern-.08em
    T\kern-.1667em\lower.7ex\hbox{E}\kern-.125emX}}

\makeatother

\begin{document}

\title{Exceptional Points of Degeneracy with Indirect Bandgap Induced By
Mixing Forward and Backward Propagating Waves}

\author{\IEEEauthorblockN{Tarek Mealy, and Filippo Capolino\textsuperscript{}}
\IEEEauthorblockA{\textit{Department of Electrical Engineering and Computer Science,
University of California, Irvine, CA 92697, USA} \\
tmealy@uci.edu, and f.capolino@uci.edu}}
\maketitle
\thispagestyle{fancy}
\begin{abstract}
We demonstrate that exceptional points of degeneracy (EPDs) are obtained
in two coupled waveguides without resorting to gain and loss. We show
the general concept that modes resulting from a proper coupling of
forward and backward waves exhibit EPDs of order two and that there
the group velocity vanishes. We verify our insight by using coupled
mode theory and also by fullwave numerical simulations of light in
a dielectric slab coupled to a grating, when one supports a forward
wave whereas the other (the grating) supports a backward wave. We
also demonstrate how to realize photonic indirect bandgaps in guiding
systems supporting a backward and a forward wave, show its relations
to the occurrence of EPDs, and offer a design procedure.
\end{abstract}


\section{Intruduction}

An EPD is a point in the parameter space of a system at which the
system's eigenvalues and eigenvectors coalesce \cite{Vishik_1960_Solution,Lancaster1964On,Seyranian1993Sensitivity,Kato1966Perturbation}.
The term exceptional point (EP) and the associated perturbation theory
where discussed in the well known Kato's book in 1966 \cite{Kato1966Perturbation}.
The phenomenon of degeneracy of both eigenvalues and eigenvectors
(polarization states), studied here, is a stronger degeneracy compared
to the traditional degeneracy of only two eigenvalues.

Non-Hermitian Hamiltonian can possess entirely real spectra when the
system obeys parity-time (PT) symmetry condition \cite{bender1998real}.
A system is said to be PT symmetric if the PT operator commutes with
the Hamiltonian \cite{bender2007making,mostafazadeh2003exact}, where
PT operator applies a parity reflection and time reversal \cite{bender1998real}.
When the time reversal operator is applied to physical systems, energy
changes from damping to growing and vice versa \cite{asadchy2020tutorial}.
Based on this simple concept, two symmetrical coupled waveguides with
balanced gain and loss satisfy PT symmetry \cite{ruschhaupt2005physical,el2007theory,guo2009observation},
where the individual application of each of the space or time reversal
would swap the gain and loss, therefore the simultaneous application
of space and time reversal operator to the system would end up with
the same system. The point separating the complex and real spectra
regimes of PT-symmetric Hamiltonians has been called exceptional point
(EP) \cite{Kato1966Perturbation}, also known as transition point.
Here, beside the mathematical aspects, we stress the role of degeneracy,
as implied also in \cite{Berry2004Physics}, hence include the 'D'
in the EPD acronym.

In this paper, we present a class of two coupled waveguides where
EPDs exist without resorting to the presence of gain and loss. By
using coupled mode theory \cite{yariv1973coupled,hardy1985coupled},
we show that two coupled waveguides, where one waveguide supports
forward propagation (i.e., where the phase and power propagate in
the same direction) whereas the other one supports backward propagation
(i.e., where the phase and power propagate in opposite direction),
experience a phase transition as in the PT-symmetric case. We show
the general conditions for modes resulting from coupling two coupled
waves to exhibit an EPD looking at both the degenerate eigenvalues
and eigenvectors. We show that the coupling of two waves, that carry
power in opposite directions, leads to an EPD and we explain how this
results in the vanishing of the group velocity of the degenerate mode.
We illustrate the concepts in a simple system made of two coupled
waveguides, i.e., a dielectric slab coupled to a grating, when one
supports a forward wave whereas the other (the grating) supports a
backward wave. Other general conditions that lead to exceptional degeneracies
of two modes in uniform waveguide were studied in \cite{mealy2020general}
using a transmission line approach. Finally, we relate the occurrence
of EPDs to the presence of a photonic indirect bandgap.

\begin{figure}
\centering{}\includegraphics[width=0.93\columnwidth]{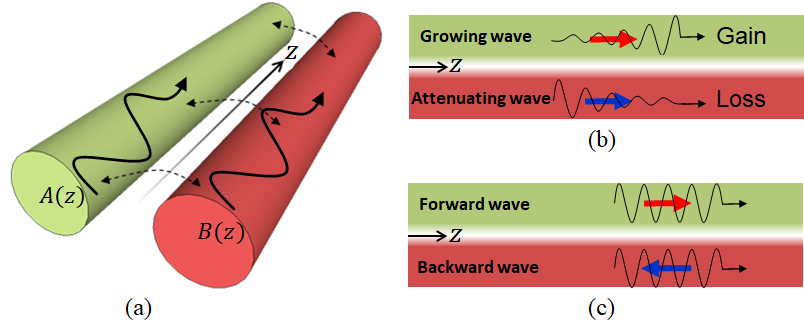}\caption{(a) Coupling between two electromagnetic waves whose complex amplitudes
are $A$ and $B$. Conditions that lead to EPDs are obtained by introducing
proper coupling between: (b) two waveguides with PT symmetry where
the two media have gain and loss supporting exponentially growing
and attenuating waves; (c) two waveguides with forward and backward
propagating waves, without resorting to PT symmetry (i.e., in this
case the waveguides do not have gain and loss). The waves with black
arrows represent the directions of propagation. The blue and red arrows
represent the directions of power flow.}
\label{fig:Coupled_modes}
\end{figure}

\section{Second order EPD by mixing two waves}

We consider two coupled electromagnetic waves as shown in Fig. \ref{fig:Coupled_modes}(a).
These two waves are described by the complex time-domain notation 

\begin{equation}
\begin{array}{c}
a(\mathbf{r},t)=A(z)f_{a}(\mathbf{\boldsymbol{\rho}})e^{i\omega t},\\
b(\mathbf{r},t)=B(z)f_{b}(\mathbf{\boldsymbol{\rho}})e^{i\omega t},
\end{array}
\end{equation}
where $A(z)$ and $B(z)$ are the complex amplitudes of waves along
the $z$ direction, $\mathbf{r}=\boldsymbol{\rho}+z\hat{\mathbf{z}}$,
and $\boldsymbol{\rho}$ is the transverese coordinate. $f_{a}(\mathbf{\boldsymbol{\rho}})$
and $f_{b}(\mathbf{\boldsymbol{\rho}})$ are the normalized modal
field profile in the transverse direction for each mode. When the
two waves are uncoupled, i..e., when their waveguides are far from
from each other, the evolution of the amplitudes along the $z$ direction
is simply descried by $dA(z)/dz=-i\beta_{a}^{\prime}A(z)$ and $dB(z)/dz=-i\beta_{b}^{\prime}B(z)$,
where $\beta_{a}^{\prime}$ and $\beta_{b}^{\prime}$ are the uncoupled
propagation constants of each wave (that is also an eigenmode of the
structure since there is no coupling). The solutions are $A(z)=A_{0}\exp(-i\beta_{a}^{\prime}z)$
and $B(z)=B_{0}\exp(-i\beta_{a}^{\prime}z)$. The power carried by
each wave in the positive $z$ direction is given by $p_{a}(z)=\pm|A(z)|^{2}$
and $p_{b}(z)=\pm|B(z)|^{2}$ where the sign depends on the type of
the wave. The sign is positive when the wave is forward, i.e., when
the power is carried in the same direction of wave propagation (i.e.,
when the phase and group velocity have the same directions). The sign
is negative when the waves backward, i.e., when the phase propagates
along the positive $z$ direction whereas the power flows in the negative
$z$ direction (i.e., when the phase and group velocity have the opposite
directions).

When coupling is introduced to those two waves, the system eigenmode
is found by solving the spatial-evolution equation that, based on
coupled mode theory \cite{yariv1973coupled,hardy1985coupled}, is
given by

\begin{equation}
\dfrac{d}{dz}\left(\begin{array}{c}
A(z)\\
B(z)
\end{array}\right)=-i\left(\begin{array}{cc}
\beta_{a} & \kappa_{ab}\\
\kappa_{ba} & \beta_{b}
\end{array}\right)\left(\begin{array}{c}
A(z)\\
B(z)
\end{array}\right),\label{eq:Coupled_DE}
\end{equation}
where $\beta_{a}$ and $\beta_{b}$ are ``perturbed'' propagation
constants for the coupled system, and $\kappa_{ab}$ and $\kappa_{ba}$
are the coupling coefficients between the two modes. The relation
between $\kappa_{ab}$ and $\kappa_{ba}$ is determined by applying
the power conservation principle. The total power carried in the coupled
structure is $p_{t}(z)=|A(z)|^{2}\pm|B(z)|^{2}$ assuming wave $A$
is forward wave, and wave $B$ to be either forward or backward, when
taking the $+$ or the $-$ sign, respectively. Thus, there are two
possible scenarios: (i) ``codirectional coupling'' when both waves
carry power at the same direction and (ii) ``contradirectional coupling''
when the two waves carry power in the opposite direction \cite{yariv1973coupled}.

When the system does not have gain and loss, conservation of energy
states that $dp_{t}(z)/dz=0$, and by using (\ref{eq:Coupled_DE}),
one finds that the constraint $\mathrm{Re}\left(AB^{*}\left(\kappa_{ba}\mp\kappa_{ab}^{*}\right)\right)=0$
should be satisfied. Therefore, we have $\kappa_{ab}=\kappa_{ba}^{*}$
in case of codirectional coupling where the two waves are forward,
and $\kappa_{ab}=-\kappa_{ba}^{*}$ in case of contradirectional coupling
where one wave is a forward and the other one is backward \cite{yariv1973coupled}.

The mixing of the two waves constitutes what is called the guiding
system's eigenmode (some call it ``supermode'') which is a weighted
sum of the individual guided waves. The eigenmode propagation constant
is determined by solving the characteristic equation of the coupled
system in (\ref{eq:Coupled_DE}) assuming the wave amplitudes to be
in the form of $[A(z),B(z)]^{T}\propto e^{-ikz}$ which yields $k^{2}-k(\beta_{a}+\beta_{b})+\left(\beta_{a}\beta_{b}-\kappa_{ab}\kappa_{ba}\right)=0.$
The characteristic equation has two solutions that are given by

\begin{equation}
\begin{array}{c}
k_{n}=\dfrac{\beta_{a}+\beta_{b}}{2}+(-1)^{n}\sqrt{\left(\dfrac{\beta_{a}-\beta_{b}}{2}\right)^{2}-\left(-1\right)^{p}\kappa^{2}},\end{array}\label{eq:CE_Sol}
\end{equation}
where $\kappa=|\kappa_{ab}|$ and the indices $n=1,2$ denote the
two modes of the coupled system. Furthermore, $p=1$ and $p=2$ represent
the case of codirectional and contradirectional coupling, respectively.
An EPD occurs when two eigenmodes coalesce, i.e., $k_{1}=k_{2}=k_{e},$
with $k_{e}=\left(\beta_{a}+\beta_{b}\right)/2$. This EPD occurs
when $\beta_{a}-\beta_{b}=2\sqrt{\left(-1\right)^{p}\kappa^{2}}$.
At an EPD, the eigenvectors must coalesce, and in this simple system
their coalescence follows from the coalescence of the eigenvalues.
Indeed, the two eigenvectors are $\left[\begin{array}{cc}
A_{n}, & B_{n}\end{array}\right]^{T}=\left[\begin{array}{cc}
1, & (k_{n}-\beta_{a})/\kappa_{ab}\end{array}\right]^{T}$and it is easy to see that they coalesce when $k_{1}=k_{2}$.

The group velocity of the eigenmode with wavenumber $k_{n}$ is determined
as (assuming $k_{n}$ to be purley real) 

\begin{equation}
v_{g,n}=\frac{1}{d_{\omega}k_{n}}=\dfrac{k_{n}-k_{e}}{2k_{n}d_{\omega}\left(\beta_{a}+\beta_{b}\right)+2d_{\omega}\left(\beta_{a}\beta_{b}+\left(-1\right)^{p}\kappa^{2}\right)}.\label{eq:Vg_Formula}
\end{equation}
where $d_{\omega}\equiv d/d\omega$ denotes the derivative with respect
to angular frequency $\omega$. It is clear from the expression that
$v_{g,1,2}=0$ when $k_{1}=k_{2}=k_{e},$i.e., exactly at the EPD.
Next, we also show what happens near the EPD. 

\begin{figure}
\begin{centering}
\centering \subfigure[]{\includegraphics[width=0.49\columnwidth]{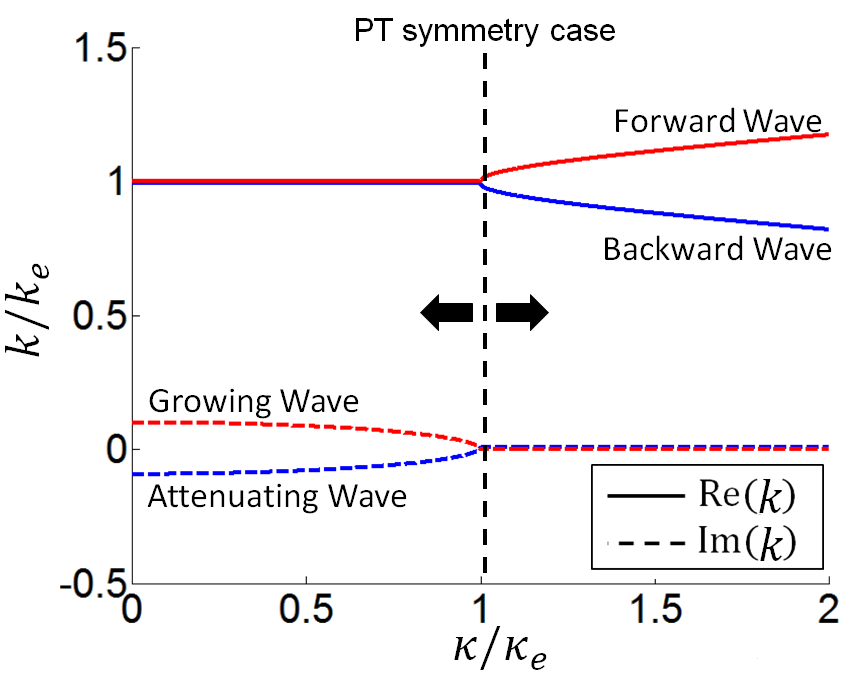}}
\subfigure[]{\includegraphics[width=0.49\columnwidth]{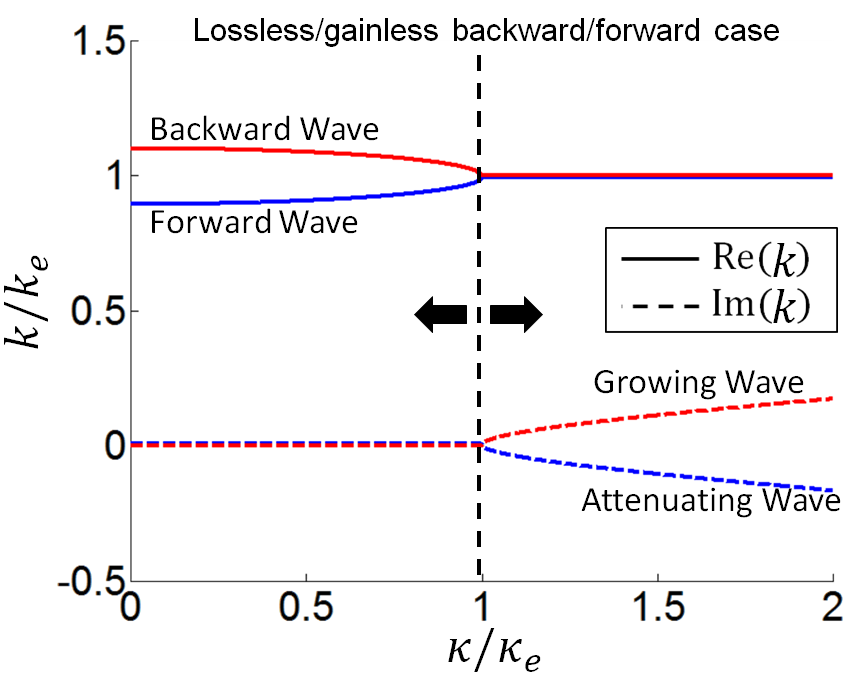}} 
\par\end{centering}
\caption{The two wavenumbers of the guiding system versus the coupling parameter
$\kappa$, showing the existence of an EPD. Two cases are examined:
(a) coalescence of modes in PT-symmetrical waveguides, (b) coalescence
of modes obtained by coupling a forward wave (phase and group velocities
have the same direction) and a backward wave (phase and group velocities
have opposite directions). Both cases exhibit an EPD, represented
by the bifurcation point.}

\centering{}\label{Fig:disp_CMT}
\end{figure}

\subsection{Codirectional coupling $(\kappa_{ab}=\kappa_{ba}^{*})$}

For the case of codirectional coupling , $p=1,$ the EPD condition
is simplified to $\beta_{a}-\beta_{b}=\pm2i\kappa$. The EPD condition
puts a constraint that the difference between the propagation constants
of the uncoupled waves has to be purely imaginary in order to exhibit
an EPD. Thus, we conclude that the EPD can never be obtained for any
value of the coupling parameter $\kappa$ in the case of a lossless/gainless
system. If we resort to a PT symmetry, as in Fig. \ref{fig:Coupled_modes}(b),
where the system has balanced gain and loss, we have $\beta_{a}=\beta_{0}+i\alpha$
and $\beta_{b}=\beta_{0}-i\alpha$, and an EPD is obtained when $\alpha=\kappa$
\cite{ruschhaupt2005physical,el2007theory,guo2009observation,Coupled_Mode2,othman2016theory},
and the degenerate wavenumber is $k_{e}=\beta_{0}$.

For this case, the two propagation constants of the coupled system
in the vicinity of the EPD are $k_{1,2}=\beta_{0}\pm\sqrt{\kappa^{2}-\alpha^{2}}$
and their derivatives are $d_{\omega}k_{1,2}=d_{\omega}\beta_{0}\pm\left(2\kappa d_{\omega}\kappa-2\alpha d_{\omega}\alpha\right)/(k_{2}-k_{1})$.
This is also illustrated by determining the eigenvector of the degenerate
eigenmode from (\ref{eq:Coupled_DE}) (for codirectional coupling
case where $p=1$) as

\begin{equation}
\begin{array}{c}
\left(\begin{array}{c}
A_{e}(z)\\
B_{e}(z)
\end{array}\right)=\left(\begin{array}{c}
1\\
-ie^{-i\mathrm{arg}(\kappa_{ab})}
\end{array}\right)e^{-ik_{e}z},\end{array}
\end{equation}
and one finds that the total power carried by the degenerate eigenmode
is $p_{t}(z)=|A_{e}(z)|^{2}-|B_{e}(z)|^{2}=0$ vanishes, in agreement
with the vanishing of the group velocity. In the vicinity of an EPD
we have $k_{1}\approx k_{2}$ and by neglecting the $d_{\omega}\beta_{0}$,
usually smaller than the other term, the group velocities of the two
modes are $v_{g,1,2}\approx\pm\left(k_{2}-k_{1}\right)/\left(2\kappa d_{\omega}\kappa-2\alpha d_{\omega}\alpha\right)$
when $\kappa>\kappa_{e}$, i.e., where the two eigenmodes are propagating
with purley real wavenumbers. Therefore we conclude that near an EPD
in a PT-symmetric guiding system, the two modes of the coupled system
are phase-synchronized ($k_{1}\approx k_{2}$), i.e., with almost
identical phase velocity, but have opposite group velocity ($v_{g,1}\approx-v_{g,2}$),
which eventually results on having a wave in the guiding structure
with vanishing group velocity when the system is exactly at the EPD.
As an example, for a system with PT symmetry where the uncoupled waveguides
have, respectively, a growing wave with $\beta_{a}=100+10i$ (1/m)
and an attenuating wave with $\beta_{b}=100-10i$ (1/m), an EPD is
obtained when the coupling parameter is $\kappa=10$ 1/m as shown
in Fig. \ref{Fig:disp_CMT}(a).

\subsection{Contradirectional coupling $(\kappa_{ab}=-\kappa_{ba}^{*})$}

We consider coupling between forward wave with wavenumber $\beta_{a}>0$
and $v_{g,a}>0$ ($v_{g,a}=d_{\omega}\beta_{a}$) and backward wave
with wavenumber $\beta_{b}>0$ and $v_{g,b}<0$ ($v_{g,b}=d_{\omega}\beta_{b}$).
The two propagating waves carry power in opposite directions and therefore
they exhibit contradirectional coupling, we use $p=2$ in Eq. (\ref{eq:CE_Sol}).
The EPD condition ($k_{1}=k_{2}$) for this case is simplified to
$\beta_{a}-\beta_{b}=\pm2\kappa$, which means that the difference
between the propagation constants should be purely real to have an
EPD, which is possible for a lossless/gainless system. Therefore,
the EPD condition for this case is satisfied through the proper design
of the coupling parameters, i.e, when $\kappa=\left|\beta_{a}-\beta_{b}\right|/2$
(we recall that $\kappa$ was defined as purely real positive). This
means that there are two possible EPD conditions, $\beta_{a}-\beta_{b}=2\kappa$
and $\beta_{a}-\beta_{b}=-2\kappa$, that may both occur when varying
frequency. At those two frequencies one has $\beta_{a}>\beta_{b}$
and $\beta_{b}>\beta_{a}$, respectively. A more detailed discussion
is provided later on when discussing the indirect bandgap.

In the \textit{vicinity} of an EPD, the two propagation constants
of the coupled system are given by Eq. (\ref{eq:CE_Sol}), and the
derivatives of the two wavenumbers with respect to the angular frequency
are

\begin{equation}
d_{\omega}k_{1,2}=\frac{1}{2}d_{\omega}\left(\beta_{a}+\beta_{b}\right)\pm\frac{1}{2}\dfrac{\left(\beta_{a}-\beta_{b}\right)d_{\omega}\left(\beta_{a}-\beta_{b}\right)-2\kappa d_{\omega}\kappa}{(k_{2}-k_{1})}.
\end{equation}

When $\kappa<\kappa_{e}$, i.e., where the two eigenmodes are propagating
with purely real wavenumbers, in the vicinity of an EPD we have $k_{1}\approx k_{2}$
and by neglecting the term $d_{\omega}\left(\beta_{a}+\beta_{b}\right)$
with respect to the second one, the group velocities of the two modes
are

\begin{equation}
v_{g,1,2}\approx\pm\frac{k_{2}-k_{1}}{\frac{1}{2}\left(\beta_{a}-\beta_{b}\right)\left(v_{g,a}^{-1}-v_{g,b}^{-1}\right)-2\kappa d_{\omega}\kappa}.
\end{equation}
Therefore, we conclude that near an EPD, the coupled forward and backward
waves are syncronised in phase, i.e., $k_{1}\approx k_{2}$ but have
nearly opposite group velocity ($v_{g,1}\approx-v_{g,2}$), which
eventually results in having a wave in the guiding structure with
vanishing group velocity when the system is exactly at an EPD.

This is also illustrated by determining the eigenvector of the degenerate
eigenmode from (\ref{eq:Coupled_DE}) (for contradirectional coupling
case where $p=2$) as

\begin{equation}
\begin{array}{c}
\left(\begin{array}{c}
A_{e}(z)\\
B_{e}(z)
\end{array}\right)=\left(\begin{array}{c}
1\\
-e^{-i\mathrm{arg}(\kappa_{ab})}
\end{array}\right)e^{-ik_{e}z},\end{array}
\end{equation}
and one finds that the total power carried by the degenerate eigenmode
is $P_{t}(z)=|A_{e}(z)|^{2}-|B_{e}(z)|^{2}=0$ vanishes, in agreement
with the vanishing of the group velocity. At the EPD, the system matrix
is not diagonalizable but rather similar to a $2\times2$ Jordan matrix.
The fields in the two-waveguide system is represented using the degenerate
and generalized eigenvectors as

\begin{equation}
\begin{array}{c}
\left(\begin{array}{c}
A(z)\\
B(z)
\end{array}\right)=\left(\begin{array}{c}
1\\
-e^{-i\mathrm{arg}(\kappa_{ab})}
\end{array}\right)\left(u_{1}-jzu_{2}\right)e^{-ik_{e}z}\ \ \ \ \ \ \ \ \ \ \ \ \ \ \\
\ \ \ \ \ \ \ \ \ \ \ \ \ \ +\left(\begin{array}{c}
1\\
ie^{-i\mathrm{arg}(\kappa_{ab})}/\kappa_{e}
\end{array}\right)u_{2}e^{-ik_{e}z}.
\end{array}
\end{equation}
 where $u_{1}$ and $u_{2}$ are proper coefficients that depend on
the system excitation and boundary conditions.

The two waveguides with contradirectional coupling are schematically
shown in Fig. \ref{fig:Coupled_modes}(c) and the dispersion diagram
is in Fig. \ref{Fig:disp_CMT} where we see a forward wave with $\beta_{a}=110$
(1/m) and backward wave with $\beta_{b}=90$ (1/m), and the EPD is
obtained at $k=k_{e}=100$ (1/m) , where $\kappa_{e}=10$ (1/m) as
shown in Fig. \ref{Fig:disp_CMT}(b).

The contradirectional coupling case can be realized is in two possible
scenarios: (i) two modes exist in two separate waveguides where the
first waveguide supports a forward wave and the second waveguide supports
a backward wave and the coupling is introduced by bringing them near
each other; (ii) two modes exist in the same waveguide having periodicity
where the one wave (e.g., the forward) has the fundamental Floquet
harmonic equal to $\beta_{1}=\beta_{0}$ and the other wave (e.g.,
the backward) has its 1st harmonic Floquet harmonic equal to $\beta_{2}=-\beta_{0}+2\pi/d$,
where $d$ is the waveguide period, and the EPD is only possible at
the band edge $\beta_{0}=\pi/d$. An example belonging to the first
scenario, where the EPD is found in two coupled dielectric slab waveguides,
is shown later on. The second scenario instead exists in conventional
periodic waveguides and it is not further considered in this paper.

\begin{figure}
\begin{centering}
\centering \subfigure[]{\includegraphics[width=0.6\columnwidth]{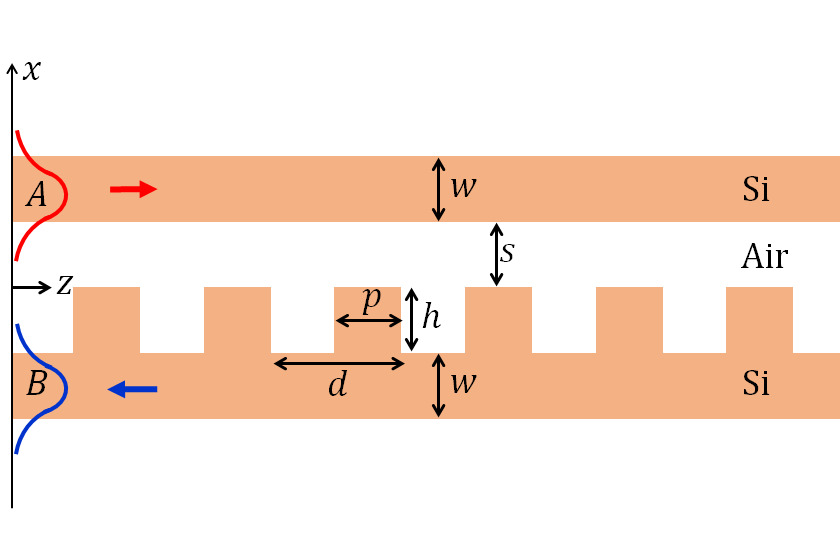}\label{Fig:Slab_WG}}
\centering
\par\end{centering}
\begin{centering}
\subfigure[]{\includegraphics[width=0.75\columnwidth]{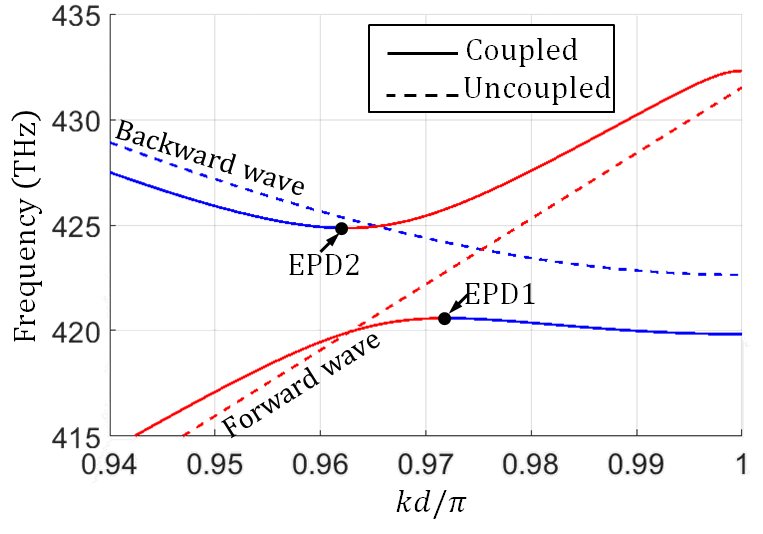}\label{Fig:Slab_disp}} 
\par\end{centering}
\caption{Example of modal EPD of order 2 between a forward wave (phase and
group velocities have the same direction) and backward wave (phase
and group velocities have opposite directions). (a) Two coupled Si
layers where the top one supports a forward wave (in red) while the
bottom one is periodically corrugated to support a backward wave (positive
phase velocity and negative group velocity, in blue). The red and
blue arrows represent the direction of power flow. (b) Dispersion
relation showing the propagating eigenmodes when the two waveguides
are uncoupled (dashed) and when the are coupled (solid). The dispersion
of modes in the coupled waveguides show the existence of two EPDs.
The blue and red colors of the curves are related to the power flow
directions. Note also that because two EPDs are found, an indirect
bandgap is present between the upper and lower branches that can be
designed ad-hoc. In the shown case we have $\Delta k_{e}\equiv k_{e2}-k_{e1}<0$.}
\end{figure}

We present an example of a guiding system that supports two waves
carrying power in opposite directions and we show that it exhibits
two EPDs. Consider the guiding system made of a Si substrate (supporting
the forward wave) coupled to a Si grating waveguide as shown in Fig.
\ref{Fig:Slab_WG}, with dimensions $w=p=h=70$ nm and $d=140$ nm.
Silcon is modeled with a refractive index $n_{\mathrm{Si}}=3.45$.
We first show in Fig. \ref{Fig:Slab_disp} the dispersion of the two
wavenumbers $\beta_{a}^{\prime}$ and $\beta_{b}^{\prime}$ of the
two uncoupled waveguides ($s\to\infty$) as red dashed (uniform waveguide)
and blue dashed (grating waveguide). The figure shows that the structures
support a forward wave where the group velocity is positive $v'_{g,a}=1/(d_{\omega}\beta_{a}^{\prime})>0$
and a backward wave where the group velocity is negative $v'_{g,b}=1/(d_{\omega}\beta_{b}^{\prime})<0$.
In the same Fig. \ref{Fig:Slab_disp}, we show the dispersion of the
two wavenumbers $k_{1}$ and $k_{2}$ of the coupled guiding system,
i.e., when the two waveguide are close to each other with a gap of
$s=70$ nm. The dispersion show the existence of two EPDs associated
to a wavenumber (momentum) displacement. The dispersion diagrams we
show are for modes with electric field polarized in the $y$ direction.
The dispersion diagrams have been found by using the finite element
method-based eigenmode solver implemented in CST Studio Suite, by
numerically simulating only one unit cell of the structure. The proposed
condition allows to locate the EPDs at band edges that are not necessarily
at the center or at the edge of a Brillioum zone, without using loss
and gain. It also shows the capability to engineer an indirect bandgap
in these simple structures.

\section{Indirect bandgap in the contradirectional case}

\begin{figure}
\begin{centering}
\includegraphics[width=0.8\columnwidth]{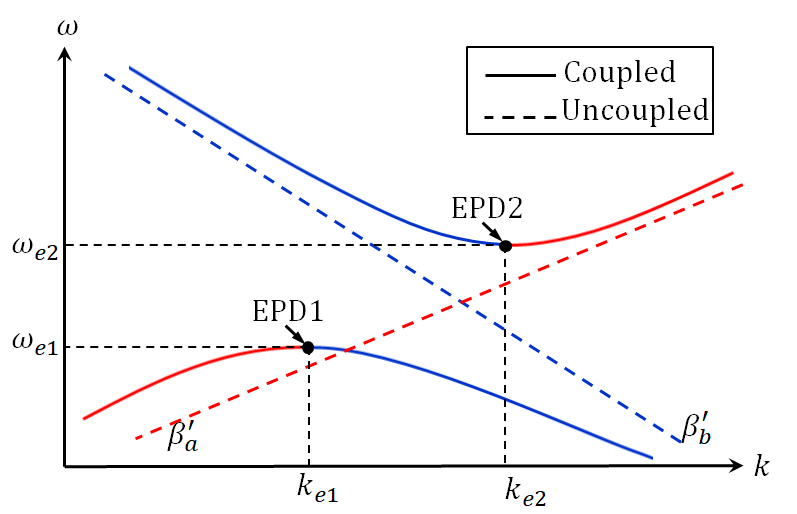}
\par\end{centering}
\caption{Schematic of a dispersion diagram showing the indirect bandgap that
results from two EPDs based on contradirectional coupling. The red-dashed
line represents the wavenumber of the forward wave $\beta_{a}^{\prime}$
whereas the blue-dahsed line the one of the backward wave $\beta_{b}^{\prime}$,
when the waveguides are uncoupled. The coupling yields the two curves
with two EPDs that are labeled as EPD1 and EPD2. The indirect bandgap
width is $\Delta\omega_{IB}$. In the shown case $\Delta k_{e}\equiv k_{e2}-k_{e1}>0$.}
\label{Fig:schematic_Indirect}
\end{figure}

In the contradirectional case where coupling occurs between a forward
and backward wave, as in Fig. \ref{fig:Coupled_modes}(c), an indirect
bandgap is possible and we show here how it is formed. Since in this
case one wave is forward and one is backward, the uncoupled propagation
constants $\beta_{a}^{\prime}$ and $\beta_{b}^{\prime}$ have opposite
slopes as schematically shown in Fig. \ref{Fig:schematic_Indirect}
(see also dashed blue and red curves in Fig. \ref{Fig:Slab_disp}),
and an analogous trend is expected for the parameters $\beta_{a}$
and $\beta_{b}$ of the coupled system. By looking at the dispersion
diagram shown in Fig. \ref{Fig:schematic_Indirect}, the red-dashed
curve is the forward wave with wavenumber $\beta_{a}^{\prime}(\omega)$
and the blue-dashed curve is the backward wave with wavenumber $\beta_{b}^{\prime}(\omega)$.
Assuming that the coupling it not so strong, one may assume that $\beta_{a}(\omega)\approx\beta_{a}^{\prime}(\omega)$
and $\beta_{b}(\omega)\approx\beta_{b}^{\prime}(\omega)$, at least
in trend, hence, in slope. The forward wave has positive slope, $v_{g,a}=d\omega/d\beta_{a}>0$,
whereas the backward wave has negative slope $v_{g,b}=d\omega/d\beta_{b}<0$,
and the dispersion curves for $\beta_{a}$ and $\beta_{b}$ versus
frequency should intersect at some frequency ($\beta_{a}=\beta_{b}$)
because one wavenumber is increasing with frequency whereas the other
is decreasing with frequency; an example with a grating is illustrated
in Fig. \ref{Fig:Slab_disp} while a schemtic is in Fig. \ref{Fig:schematic_Indirect}.
Let us approximate the dispersion curves locally, in the frequency
range of interest, as straight lines, i.e., $\beta_{a}(\omega)\approx a+v_{g,a}^{-1}\omega$
and $\beta_{b}(\omega)\approx b+v_{g,b}^{-1}\omega$, where now $v_{g,a}$
and $v_{g,b}$ are assumed to have the local fixed value. Furthermore,
assuming, for simplicity, that $\kappa$ is constant within the frequency
range of interest, one finds that EPDs occurs at two angular frequencies
$\omega_{e1}$ and $\omega_{e2}$ such that $\beta_{a}(\omega_{e1})-\beta_{b}(\omega_{e1})=-2\kappa$
and $\beta_{a}(\omega_{e2})-\beta_{b}(\omega_{e2})=2\kappa$, where
$\omega_{e1}<\omega_{e2}$. Subtracting the previous two conditions
leads to the indirect bandgap determination 
\begin{equation}
\Delta\omega_{IB}\equiv\omega_{e2}-\omega_{e1}\approx\frac{4\kappa}{v_{g,a}^{-1}-v_{g,b}^{-1}}.\label{eq:BandgapWidth}
\end{equation}

Note that $v_{g,a}^{-1}-v_{g,b}^{-1}>0$, hence $\Delta\omega_{IB}>0$.
The bandgap width can be controlled by the slope of the two parameters
$\beta_{a}$ and $\beta_{b}$; indeed, when $v_{g,a}^{-1}\approx-v_{g,b}^{-1}$,
the denominator of (\ref{eq:BandgapWidth}) is small and the bandgap
is very wide, viceversa, the bandgap is narrow when $v_{g,a}^{-1}$
is very different from $-v_{g,b}^{-1}$. If we consider the dispersion
of the coupling term $\kappa$, a more complicated picture may arise
that could be determined by the reasoning just provided.

The degenerate wavenumbers at the two EPDs are $k_{e,1}=\left(\beta_{a}(\omega_{e1})+\beta_{b}(\omega_{e1})\right)/2$
and $k_{e,2}=\left(\beta_{a}(\omega_{e2})+\beta_{b}(\omega_{e2})\right)/2$.
Using the linear approximation formulas for the wavenumbers $\beta_{a}(\omega)$
and $\beta_{a}(\omega)$, one finds that $k_{e1}\approx[a+b+\left(v_{g,a}^{-1}+v_{g,b}^{-1}\right)\omega_{e1}]/2$
and $k_{e2}\approx[a+b+\left(v_{g,a}^{-1}+v_{g,b}^{-1}\right)\omega_{e2}]/2$
. The difference between the two degenerate wavenumbers is

\begin{equation}
\Delta k_{e}\equiv k_{e2}-k_{e1}\approx\frac{v_{g,a}^{-1}+v_{g,b}^{-1}}{2}\Delta\omega_{IB}.\label{eq:Delta_ke_Eq_CMT}
\end{equation}
Therefore, it is necessary that $v_{g,a}^{-1}+v_{g,b}^{-1}\neq0$
in order to have indirect bandgap. When $|v_{g,a}^{-1}|>|v_{g,b}^{-1}|$
we get $k_{e2}>k_{e1},$hence $\Delta k_{e}>0$, i.e., the EPD that
occurs at the smaller frequency $\omega_{e1}$ occurs also at the
smaller degenerate wavenumber $k_{e1}$ and this condition is depicted
in Fig. \ref{Fig:schematic_Indirect}. When $|v_{g,a}^{-1}|<|v_{g,b}^{-1}|$,
we get $k_{e1}>k_{e2},$ hence $\Delta k_{e}<0$, i.e., the EPD that
occurs at the smaller frequency $\omega_{e1}$ occurs at larger degenerate
wavenumber $k_{e1}$ and this condition is depicted in Fig. \ref{Fig:Slab_disp}.
Indeed, by looking at Fig. \ref{Fig:Slab_disp}, one finds by naked
eye that $|v_{g,a}|>|v_{g,b}|$ , (absolute change with frequency
of dashed red curve is higher than the one of the dashed blue one),
therefore $|v_{g,a}^{-1}|<|v_{g,b}^{-1}|$ resulting in $k_{e1}>k_{e2}$
according to (\ref{eq:Delta_ke_Eq_CMT}) and the EPD at lower frequency,
around 420 THz, occurs at higher degenerate wavenumber of $k_{e1}=0.972\,\pi/d$
.

\section{Conclusion}

We have demonstrated that EPDs are not only obtained in PT-symmetric
waveguides but they are also obtained in two lossless and gainless
waveguides when they support forward and backward waves that are properly
coupled. We have shown a simple system that supports this condition
made of a grating (supporting a backward guided mode) coupled to a
dielectric layer (supporting the forward mode). We have elaborated
that the scheme discussed here exhibits a photonic indirect bandgap
that can be controlled by changing the group velocities of the forward
and backward modes along with the coupling coefficient. Two conditions
may occur: the bandgap is associated either to a positive or negative
momentum difference between the two energy levels. The finding in
this paper can be useful to design systems with EPDs whose use is
of growing importance for enhancing light-matter interactions and
nonlinear photonic phenomena. 

\section*{Acknowledgment}

This material is based on work supported by the Air Force Office of
Scientific Research award number FA9550-18-1-0355, and by the National
Science Foundation under award NSF ECCS-1711975. The authors are thankful
to DS SIMULIA for providing CST Studio Suite that was instrumental
in this study. The authors thank Dr. Hamidreza Kazemi, Dr. Mohamed
Yehia Nada, and Dr. Ahmed Abdelshafy, UC Irvine, for fruitful discussions
on coupled mode theory.

\bibliographystyle{ieeetr}
\bibliography{EPD_General_Cond_Ref}

\end{document}